\documentstyle[11pt,epsf]{article}
\newtheorem{theorem}{Theorem}

\newcommand {\dfn} {\stackrel{\Delta} {=}}
\newcommand {\exe} {\stackrel{\cdot} {=}}
\newcommand {\lexe} {\stackrel{\cdot} {\le}}

\newcommand {\bx} {\mbox{\boldmath $x$}}
\newcommand {\by} {\mbox{\boldmath $y$}}

\newcommand {\bE} {\mbox{\boldmath $E$}}

\newcommand{\calB}{{\cal B}}
\newcommand{\calC}{{\cal C}}

\newcommand{\calG}{{\cal G}}

\newcommand{\calT}{{\cal T}}

\newcommand{\calX}{{\cal X}}
\newcommand{\calY}{{\cal Y}}

\begin{document}
\thispagestyle{empty}
\title{Correction to ``The Generalized Stochastic Likelihood Decoder: Random
Coding and Expurgated Bounds''}
\author{Neri Merhav}
\date{}
\maketitle

\begin{center}
The Andrew \& Erna Viterbi Faculty of Electrical Engineering\\
Technion - Israel Institute of Technology \\
Technion City, Haifa 32000, ISRAEL \\
E--mail: {\tt merhav@ee.technion.ac.il}\\
\end{center}
\vspace{1.5\baselineskip}
\setlength{\baselineskip}{1.5\baselineskip}

\begin{abstract}
This purpose of this letter is to handle a gap
that was found in the proof of Theorem 2 in the
paper ``The generalized stochastic likelihood decoder: random coding and
expurgated bounds.''
\end{abstract}

\section{Introduction}

In a recent article \cite{p187}, random coding error exponents and expurgated
exponents were analyzed for the generalized likelihood
decoder (GLD), where the decoded message is randomly selected under a
probability distribution
that is proportional to a general exponential function of the empirical joint
distribution of the codeword and the channel output vectors.
In Section V of \cite{p187}, Theorem 2 provides an expurgated
exponent which is applicable to this decoder (and hence also to the optimal
maximum likelihood decoder). The proof of that theorem is based on two steps of a
certain expurgation procedure. Nir Weinberger has brought to my attention that
there is a certain gap in that proof, as the second expurgation step might
interfere with the first step (more details will follow in Section 2 of this
letter). The purpose of this letter is to provide an alternative proof to the
above mentioned theorem.

\section{Setup and Background}

Consider a discrete memoryless channel (DMC), 
designated by a matrix of single--letter input--output
transition probabilities $\{W(y|x),~x\in\calX,~y\in\calY\}$. Here the channel input
symbol $x$ takes on values in a finite input alphabet $\calX$, and the
channel output symbol $y$ takes on values in a finite output alphabet $\calY$.
When the channel is fed by a vector
$\bx=(x_1,\ldots,x_n)\in\calX^n$, it outputs a vector
$\by=(y_1,\ldots,y_n)\in\calY^n$ according to
\begin{equation}
\label{channel}
W(\by|\bx)=\prod_{t=1}^n W(y_t|x_t).
\end{equation}
A code $\calC_n\subseteq\calX^n$ is a collection of $M=e^{nR}$ channel input
vectors,
$\{\bx_0,\bx_1,\ldots,\bx_{M-1}\}$, $R$ being the coding rate in nats per
channel
use. It is assumed that all messages, $m=0,1,\ldots.M-1$, are equally likely.

As is very common in the information theory literature, we consider
the random coding regime.
The random coding ensemble considered
is the ensemble of constant composition codes, where each
codeword is drawn independently under the uniform distribution within a given
type class $\calT(Q_X)$, i.e., the set of all
vectors in $\calX^n$ whose empirical distribution is given by $Q_X$. 
Once the code has been randomly selected, it is
revealed to
both the encoder and the decoder.

When the transmitter wishes to convey a message $m$, it transmits the
corresponding code-vector $\bx_m$ via the channel, which in turn,
stochastically maps it into an
$n$--vector $\by$ according to (\ref{channel}).
Upon receiving $\by$,
the stochastic generalized likelihood decoder randomly selects the
estimated message $\hat{m}$ according to a generalized version of 
the induced posterior distribution
of the transmitted
message, i.e.,
\begin{eqnarray}
\label{posterior}
\mbox{Pr}\{\hat{m}=m_0|\by\}=
\frac{\exp\{ng(\hat{P}_{\bx_{m_0}\by})\}}{\sum_{m=0}^{M-1}
\exp\{ng(\hat{P}_{\bx_m\by})\}}, 
\end{eqnarray}
where $\hat{P}_{\bx_m\by}$ is the empirical joint distribution induced
by $(\bx_m,\by)$ and $g(\cdot)$ is an arbitrary continuous function.
For example,
\begin{equation}
\label{OLD}
g(\hat{P}_{\bx_m\by})=\sum_{x,y}\hat{P}_{\bx_m\by}(x,y)\ln W(y|x)
\end{equation}
corresponds to the ordinary likelihood decoder, where (\ref{posterior}) is
the correct underlying posterior probability of message $m_0$. This framework also
allows additional important stochastic decoders, where $g$ corresponds to a
mismatched metric $\tilde{W}$ or to the empirical mutual information, as
discussed in \cite{p187}.

As mentioned above, in Section V of \cite{p187}, an expurgated
error exponent is derived. Specifically, 
letting $Q_{XY}$ denote a generic joint distribution over $\calX\times\calY$,
and letting $I_Q(X;Y)$ denote the mutual information
induced by $Q_{XY}$,
we define the following.
Let
\begin{equation}
\alpha(R,Q_Y)=\sup_{\{Q_{X|Y}:~I_Q(X;Y)\le R\}}[g(Q_{XY})-I_Q(X;Y)]+R,
\end{equation}
and
\begin{eqnarray}
\Gamma(Q_{XX'},R)
&=&\inf_{Q_{Y|XX'}}\left\{D(Q_{Y|X}\|W|Q_X)+
I_Q(X';Y|X)+\right.\nonumber\\
& &\left.[\max\{g(Q_{XY}),\alpha(R,Q_Y)\}-
g(Q_{X'Y})]_+\right\}\\
&\equiv&\inf_{Q_{Y|XX'}}\left\{\bE_Q\log[1/W(Y|X)]-H(Y|X,X')+\right.\nonumber\\
& &\left.[\max\{g(Q_{XY}),\alpha(R,Q_Y)\}-g(Q_{X'Y})]_+\right\},
\end{eqnarray}
where $D(Q_{Y|X}\|W|Q_X)$ is defined in the usual manner (see also
\cite{p187}).
The main result in \cite[Section V]{p187} is the following:

\begin{theorem}
There exists a sequence of constant composition codes,
$\{\calC_n,~n=1,2,\ldots\}$,
with composition $Q_X$, such that
\begin{equation}
\liminf_{n\to\infty}\left[-\frac{\log P_{\mbox{\tiny
e}|m}(\calC_n)}{n}\right]\ge E_{\mbox{\tiny ex}}^{\mbox{\tiny gld}}(R,Q_X),
\end{equation}
where
\begin{equation}
\label{ckmstyle}
E_{\mbox{\tiny ex}}^{\mbox{\tiny gld}}(R,Q_X)
=\inf[\Gamma(Q_{XX'},R)+I_Q(X;X')]-R,
\end{equation}
where the infimum is over all joint distributions $\{Q_{XX'}\}$ such that
$I_Q(X;X')\le R$ and $Q_{X'}=Q_X$.
\end{theorem}

The proof in \cite{p187} contains two main steps of expurgation. In the first step, we
confine attention to the subset of constant composition codes $\{\calC_n\}$ with the property
\begin{equation}
\label{good}
\sum_{m'\ne m}\exp\{ng(\hat{P}_{\bx_{m'}\by})\}\ge
\exp\{n\alpha(R-\epsilon,\hat{P}_{\by})\}~~~\forall~m,\by
\end{equation}
where $\epsilon > 0$ is arbitrarily small. It is proved in \cite[Appendix
B]{p187} that the vast majority of
constant composition codes satisfy (\ref{good}) for large $n$. In the second expurgation
step (see \cite[Appendix C]{p187}), at most $M\cdot
(n+1)^{|\calX|^2}e^{-n\epsilon/2}$ ``bad''
codewords are eliminated from the codebook in order to guarantee the
desired maximum error probability performance for the remaining part of the
code. 

The gap in the proof of \cite[Theorem 2]{p187} 
is in the following point: after the second expurgation step, it is no
longer guaranteed that eq.\ (\ref{good}) still holds for every $m$ and $\by$, 
since the summation on the left--hand side of (\ref{good}) is now
taken over a smaller number of codewords.

Fortunately enough, Theorem 2 of \cite{p187} is still correct 
(as will be proved in the next section in a completely different manner) at least when
$g(Q_{XY})$ is an affine functional of $Q_{XY}$, which is the case of the
ordinary matched/mismatched stochastic likelihood decoder (\ref{OLD}) with or
without a ``temperature'' parameter (see the discussion around eqs.\ (5)--(7) of
\cite{p187}). This affinity
assumption is used only at the last step of our derivation below. Thus, when
$g(Q_{XY})$ is not affine, one merely backs off from the last step of the
derivation, and considers the second to the last expression
as the formula of the expurgated exponent.

\section{Corrected Proof of \cite[Theorem 2]{p187}}

Assuming that message $m$ was
transmitted, the probability of error of the GLD, for a given code $\calC_n$, is given by
\begin{equation}
P_{\mbox{\tiny e}|m}(\calC_n)
=\sum_{m^\prime\ne m}\sum_{\by}W(\by|\bx_m)\frac{
\exp\{ng(\hat{P}_{\bx_{m^\prime}\by})\}}
{\exp\{ng(\hat{P}_{\bx_m\by})\}+\sum_{m^\prime\ne m}
\exp\{ng(\hat{P}_{\bx_{m^\prime}\by})\}}
\end{equation}
and so, for $\rho\ge 1$,
\begin{equation}
[P_{\mbox{\tiny e}|m}(\calC_n)]^{1/\rho}
\le\sum_{m^\prime\ne m}\left[\sum_{\by}W(\by|\bx_m)
\frac{\exp\{ng(\hat{P}_{\bx_{m^\prime}\by})\}}
{\exp\{ng(\hat{P}_{\bx_m\by})\}+\sum_{m^\prime\ne m}
\exp\{ng(\hat{P}_{\bx_{m^\prime}\by})\}}\right]^{1/\rho},
\end{equation}
where we have used the inequality $(\sum_i a_i)^s\le\sum_i a_i^s$, which holds
whenever $s\le 1$ and $a_i\ge 0$ for all $i$ \cite[Exercise
4.15(f)]{Gallager68}.
Let $\calG_\epsilon=\calB_\epsilon^c$ be defined as in \cite{p187}, 
that is, the set of codes for which (\ref{good}) holds, and
consider the fact (proved in Appendix B therein), that $\mbox{Pr}\{\calB_\epsilon\}\le
\exp(-e^{n\epsilon}+n\epsilon+1)$.
We now
take the expectation over the randomness of the (incorrect part of the)
codebook,
$\calC_n^m=\calC_n\setminus\{\bx_m\}$ (where all wrong codewords are drawn from 
a given type $Q_X$), except $\bx_m$, which is kept fixed
for now. When dealing with the pairwise error probability from
$m$ to $m'$, we do this in two steps: we first average over all codewords
except $\bx_m$ and $\bx_{m'}$, and then average over the randomness of $\bx_{m'}$.
\begin{eqnarray}
& & \bE\left\{[P_{\mbox{\tiny e}|m}(\calC_n)]^{1/\rho}\bigg|\bx_m\right\}\nonumber\\
&\le&\sum_{m^\prime\ne m}\sum_{\calC_n^m}P(\calC_n^m)\left[\sum_{\by}W(\by|\bx_m)
\frac{\exp\{ng(\hat{P}_{\bx_{m^\prime}\by})\}}
{\exp\{ng(\hat{P}_{\bx_m\by})\}+\sum_{m^\prime\ne m}
\exp\{ng(\hat{P}_{\bx_{m^\prime}\by})\}}\right]^{1/\rho}\nonumber\\
&=&\sum_{m^\prime\ne m}\sum_{\calC_n^m\in\calG_\epsilon}
P(\calC_n^m)\left[\sum_{\by}W(\by|\bx_m)
\frac{\exp\{ng(\hat{P}_{\bx_{m^\prime}\by})\}}
{\exp\{ng(\hat{P}_{\bx_m\by})\}+\sum_{m^\prime\ne m}
\exp\{ng(\hat{P}_{\bx_{m^\prime}\by})\}}\right]^{1/\rho}+\nonumber\\
& &\sum_{m^\prime\ne m}\sum_{\calC_n^m\in\calB_\epsilon}P(\calC_n^m)
\left[\sum_{\by}W(\by|\bx_m)
\frac{\exp\{ng(\hat{P}_{\bx_{m^\prime}\by})\}}
{\exp\{ng(\hat{P}_{\bx_m\by})\}+\sum_{m^\prime\ne m}
\exp\{ng(\hat{P}_{\bx_{m^\prime}\by})\}}\right]^{1/\rho}\nonumber\\
&\le&\sum_{m^\prime\ne m}\sum_{\calC_n^m\in\calG_\epsilon}
P(\calC_n^m)\left[\sum_{\by}W(\by|\bx_m)
\cdot\min\left\{1,\frac{\exp\{ng(\hat{P}_{\bx_{m^\prime}\by})\}}
{\exp\{ng(\hat{P}_{\bx_m\by})\}+\exp\{n\alpha(R-\epsilon,\hat{P}_{\by})\}
}\right\}\right]^{1/\rho}+\nonumber\\
& &\sum_{m^\prime\ne m}\sum_{\calC_n^m\in\calB_\epsilon}P(\calC_n^m)\cdot 1^{1/\rho}
\nonumber\\
&\le&\sum_{m^\prime\ne m}\sum_{\calC_n^m}P(\calC_n^m)\left[\sum_{\by}W(\by|\bx_m)
\cdot\min\left\{1,\frac{\exp\{ng(\hat{P}_{\bx_{m^\prime}\by})\}}
{\exp\{ng(\hat{P}_{\bx_m\by})\}+\exp\{n\alpha(R-\epsilon,\hat{P}_{\by})\}
}\right\}\right]^{1/\rho}+\nonumber\\
& &e^{nR}\cdot\exp(-e^{n\epsilon}+n\epsilon+1)
\nonumber\\
&\lexe&\sum_{m^\prime\ne m}\bE\left(\left[\sum_{\by}W(\by|\bx_m)
\cdot\min\left\{1,\frac{\exp\{ng(\hat{P}_{\bx_{m^\prime}\by})\}}
{\exp\{ng(\hat{P}_{\bx_m\by})\}+\exp\{n\alpha(R-\epsilon,\hat{P}_{\by})\}
}\right\}\right]^{1/\rho}\bigg|\bx_m\right)\nonumber\\
&\exe&\sum_{m^\prime\ne
m}\bE\left\{\exp[-n\Gamma(\hat{P}_{\bx_m\bx_m'})/\rho]\bigg|\bx_m\right\}\nonumber\\
&\exe&\sum_{Q_{X'|X}}\bE\{N_m(Q_{X'|X})|\bx_m\}\exp\{-n\Gamma(Q_{XX'})/\rho\}\nonumber\\
&\exe&\max_{Q_{X'|X}}\exp\{n[R-I_Q(X;X')]\}\cdot
\exp\{-n\Gamma(Q_{XX'})/\rho\}\nonumber\\
&=&\exp\left\{-n\min_{Q_{X'|X}}[\Gamma(Q_{XX'})/\rho+I_Q(X;X')-R]\right\},
\end{eqnarray}
where $I_Q(X;X')$ is the mutual information induced by $Q_{XX'}$ and
$N_m(Q_{X'|X})=|\calT(Q_{X'|X}|\bx_m)\cap\calC_m|$, $\calT(Q_{X'|X}|\bx_m)$
being the conditional type class pertaining to $Q_{X'|X}$ given $\bx_m$.
Since this bound is
independent of $\bx_m$, it also holds for the unconditional expectation,
$\bE[P_{e|m}(\calC_n)]^{1/\rho}$. Now, for a given code $\calC_n$, 
index the message $\{m\}$ according to
decreasing order of $\{P_{e|m}(\calC_n)\}$. Then,
\begin{equation}
\frac{1}{M}\sum_{m=1}^M[P_{e|m}(\calC_n)]^{1/\rho}\ge
\frac{1}{M}\sum_{m=1}^{M/2}[P_{e|m}(\calC_n)]^{1/\rho}\ge
\frac{1}{M}\cdot\frac{M}{2}[P_{e|M/2}(\calC_n)]^{1/\rho}=
\frac{1}{2}\cdot[\max_mP_{e|m}(\calC_n')]^{1/\rho},
\end{equation}
where $\calC_n'$ is the good half of $\calC_n$. Thus,
\begin{eqnarray}
\bE\left\{[\max_m P_{e|m}(\calC_n')]^{1/\rho}\right\}
&\le& 2
\bE\left\{\frac{1}{M}\sum_{m=1}^M
P_{e|m}(\calC_n)^{1/\rho}\right\}\nonumber\\
&\lexe&\exp\left\{-n\min_{Q_{X'|X}}[\Gamma(Q_{XX'})/\rho+I_Q(X;X')-R]\right\}
\end{eqnarray}
which means that there exists a code of size $M/2$ with
\begin{equation}
[\max_mP_{e|m}(\calC_n')]^{1/\rho}\le 
\exp\left\{-n\min_{Q_{X'|X}}[\Gamma(Q_{XX'})/\rho+I_Q(X;X')-R]\right\},
\end{equation}
or equivalently,
\begin{equation}
\max_m P_{e|m}(\calC_n')\le 
\exp\left(-n\min_{Q_{X'|X}}\{\Gamma(Q_{XX'})+\rho[I_Q(X;X')-R]\}\right),
\end{equation}
and since this holds for every $\rho\ge 1$, we have
\begin{equation}
\max_m P_{e|m}(\calC_n')\le 
\exp\left(-n\sup_{\rho\ge 1}\min_{Q_{X'|X}}\{\Gamma(Q_{XX'})+\rho[I_Q(X;X')-R]\}\right).
\end{equation}
Now, consider the exponent,
\begin{eqnarray}
\label{maxmin}
E_{\mbox{\tiny ex}}(R,Q_X)&\dfn& \sup_{\rho\ge
1}\min_{Q_{X'|X}}\{\Gamma(Q_{XX'})+\rho[I_Q(X;X')-R]\}\\
&=&\sup_{\rho\ge
0}\min_{Q_{XX'}}\{\Gamma(Q_{XX'})+I_Q(X;X')-R+\rho[I_Q(X;X')-R]\},
\end{eqnarray}
where the marginals of $Q_{XX'}$ are constrained to the given fixed
composition, $Q_X$.
Using the definitions in \cite{p187},
\begin{eqnarray}
\Gamma(Q_{XX'})+I_Q(X;X')&=&\inf_{Q_{Y|XX'}}\left\{-\bE_Q\ln
W(Y|X)-H(Y|X,X')+\right.\nonumber\\
& &\left.I_Q(X;X')+[\max\{g(Q_{XY}),\alpha(R,Q_Y)\}-g(Q_{X'Y})]_+\right\}\nonumber\\
&=&\inf_{Q_{Y|XX'}}\left\{-\bE_Q\ln[W(Y|X)Q(X)Q(X')]-H_Q(X,X',Y)+\right.\nonumber\\
& &\left.+[\max\{g(Q_{XY}),\alpha(R,Q_Y)\}-g(Q_{X'Y})]_+\right\},
\end{eqnarray}
thus,
\begin{eqnarray}
& &\min_{Q_{XX'}}\{\Gamma(Q_{XX'})+I_Q(X;X')-R+\rho[I_Q(X;X')-R]\}\nonumber\\
&=&\min_{Q_{XX'Y}}\left\{-\bE_Q\ln[W(Y|X)Q(X)Q(X')]-H_Q(X,X',Y)+\right.\nonumber\\
& &\left.\rho[I_Q(X;X')-R]+
[\max\{g(Q_{XY}),\alpha(R,Q_Y)\}-g(Q_{X'Y})]_+\right\}.
\end{eqnarray}
Now, the first term on the right--most side 
is linear (and hence convex) in $Q_{XX'Y}$ since $Q_X$ is fixed,
the second term is convex, and the third term is convex for a given $Q_X$. As
for the fourth term, it is convex at least in the case where $g$ is affine in $Q$
(e.g., matched/mismatched likelihood metric with/without a temperature
parameter) because the function $f(x)=[x]_+$ is monotonic and convex and we
argue that $\alpha(R,Q_Y)$ is also convex since it is given by the supremum
over a family of convex functions of $Q_Y$ (as $g$ is linear and $-I_Q(X;X')$
is convex in $Q_Y$ for a given $Q_{X|Y}$). The maximum between two convex
functions is convex.
Since the objective is affine (and hence concave) in $\rho$,
we can interchange the minimization
and the maximization to obtain,
\begin{eqnarray}
E_{\mbox{\tiny
ex}}(R,Q_X)&=&\inf_{Q_{XX'}}\left\{\Gamma(Q_{XX'})+I_Q(X;X')-R+\sup_{\rho\ge
0}\rho[I_Q(X;X')-R]\right\}\nonumber\\
&=&\inf_{\{Q_{XX'}:~I_Q(X;X')\le R\}}[\Gamma(Q_{XX'})+I_Q(X;X')-R]\nonumber\\
&=&E_{\mbox{\tiny ex}}^{\mbox{\tiny gld}}(R,Q_X).
\end{eqnarray}
If the supremum and the
minimum cannot be interchanged, then, of course, the formula of the expurgated exponent
remains as in (\ref{maxmin}).

\section*{Acknowledgment}

I would like to thank Dr.\ Nir Weinberger for drawing my attention to the gap
in the proof of Theorem 2 in \cite{p187}.

\end{document}